\documentclass[apj]{emulateapj}
\usepackage{apjfonts,mathptmx}
\begin{document}

\def\psr{PSR\,J0737$-$3039}
\def\psra{PSR\,A}
\def\psrb{PSR\,B}
\def\xmm {\emph{XMM-Newton}}
\def\cha {\emph{Chandra}}
\def\flux {\mbox{erg cm$^{-2}$ s$^{-1}$}}
\def\lum {\mbox{erg s$^{-1}$}}
\def\nh {$N_{\rm H}$}

\title{\psr: Interacting Pulsars in X-rays\altaffilmark{1}}

\shorttitle{\psr: Interacting Pulsars in X-rays}
\shortauthors{A.~Pellizzoni et al.}
\journalinfo{The Astrophysical Journal, in press}
\submitted{Received 2007 December 11; accepted 2008 January 31}
%
\author{A.~Pellizzoni, A.~Tiengo, A.~De~Luca,\altaffilmark{2,3} 
P.~Esposito,\altaffilmark{3,4} and S.~Mereghetti}
\affil{INAF - Istituto di Astrofisica Spaziale e Fisica Cosmica - Milano \\
via E.~Bassini 15, 20133 Milano, Italy}
\email{alberto@iasf-milano.inaf.it}
\altaffiltext{1}{Based on observations obtained with \xmm, an ESA science mission with instruments and contributions directly funded by ESA Member States and NASA}
\altaffiltext{2}{IUSS - Istituto Universitario di Studi Superiori, viale Lungo 
Ticino Sforza 56, 27100 Pavia, Italy}
\altaffiltext{3}{Universit\`a degli Studi di Pavia, Dipartimento di Fisica 
Nucleare e Teorica, via A.~Bassi 6, 27100 Pavia, Italy}
\altaffiltext{4}{INFN - Istituto Nazionale di Fisica Nucleare, Sezione di Pavia, via A.~Bassi 6, 27100 Pavia, Italy}
\begin{abstract}

We present the results of a $\sim$230 ks long X-ray observation of the 
relativistic double-pulsar system \psr\ obtained with the \xmm\ satellite in 
2006 October. We confirm the detection in X-rays of pulsed emission from \psr A (\psra),
mostly ascribed to a soft non thermal power law component ($\Gamma\sim3.3$) 
with a 0.2--3 keV luminosity of $\sim$$1.9\times10^{30}$ \lum\ (assuming a 
distance of 500 pc). For the first time, pulsed X-ray emission from \psr B (\psrb) is 
also detected in part of the orbit. This emission, consistent with thermal 
radiation with temperature $k_BT\simeq30$ eV and a bolometric luminosity of 
$\sim$$10^{32}$ \lum, is likely powered by heating of \psrb's surface caused 
by \psra's wind. A hotter ($\sim$130 eV) and fainter ($\sim$$5\times10^{29}$ 
\lum) thermal component, probably originating from back-falling particles 
heating polar caps of either \psra\ or \psrb\ is also required by the data. 
No signs of X-ray emission from a bow shock between \psra's wind and the 
interstellar medium or \psrb's magnetosphere are present. The upper limit on 
the luminosity of such a shock component ($\sim$$10^{29}$ \lum) constrains the 
wind magnetization parameter $\sigma_M$ of \psra\ to values greater than 1.

\end{abstract}


\section{Introduction}

The short-period, double--radio pulsar system \psr\ \citep{burgay03,lyne04}, 
besides being of paramount interest as a probe for theories of strong field 
gravity \citep{kramer06}, represents a unique laboratory for studies in 
several fields, ranging from the equation of state of super-dense matter to
magneto-hydrodynamics. The system is observed nearly edge-on and consists of a 
fast, recycled radio pulsar (\psr A [\psra], period \mbox{$P=22.7$ ms}) orbiting its 
slower companion (\psr B [\psrb], \mbox{$P=2.77$ s}) with an orbital period of only 2.4 
hr. Radio observations of the two pulsars permit one to derive a wealth of
information which is not available for other neutron stars. These obviously 
include the neutron star masses and all the other geometrical and dynamical 
parameters of the system, but also information on the structure and physical 
properties of the two magnetospheres. The double-pulsar is rich in 
observational phenomena, including a short radio eclipse of A by B and orbital
modulation of the radio flux of B due to the influence of A \citep{lyne04}. The
individual pulses from B show drifting features due to the impact of the 
low-frequency electromagnetic wave in the relativistic wind from A 
\citep{mclaughlin04b}, while the eclipse of A is modulated at half the 
rotational period of B \citep{mclaughlin04a}.
High-energy observations are an important complement to these studies, in 
particular for what concerns the physics of the magnetospheric emission and 
dissipative shocks in the close environment of the two neutron stars.

X-rays could be pulsed magnetospheric or thermal emission from pulsar A, as 
seen for several other recycled pulsars \citep{zavlin02}, but they could also 
originate in the colliding winds of A and B \citep{lyutikov04}. Because of the 
interaction of the relativistic wind of the fast spinning pulsar A
($\dot{E}^{\rm{A}}_{\rm{rot}}=5.8\times10^{33}$ \lum) with the magnetosphere of
its much less energetic companion
($\dot{E}^{\rm{B}}_{\rm{rot}}=1.6\times10^{30}$ \lum), the formation of a 
bow shock, likely  emitting at high energies, is expected. The predicted 
fluxes are roughly comparable to those observed for magnetospheric and/or 
surface emission from the pulsars. A termination shock between the two pulsars 
can probe the properties of a pulsar's relativistic wind at a smaller distance
from the central engine than ever studied before. In addition, detection of an
orbital phase dependence in the X-ray emission might be expected 
\citep{lyutikov04,arons93}. Such variability could constrain the geometry of 
the emission site, thus providing new insights into the wind physics close to 
the pulsar. Alternatively, most of the high-energy emission could arise from 
the shock generated when one or both the pulsar winds interact with the
interstellar medium \citep{granot04}.

The first X-ray observation of \psr, a short (10 ks) \cha\ pointing
\citep{mclaughlin04} yielded only $\sim$80  photons. The  low X-ray luminosity
(L$_{\rm{X}}=2\times10^{30} d_{0.5}^2$ \lum, where we indicate with 
$d_N$ the distance in units of $N$ kpc) corresponds roughly to the entire 
spin-down luminosity of the slow pulsar B and  to only a small fraction of that
of pulsar A. The \cha\ data of this pioneer observation could only poorly
constrain the source spectrum, which appeared quite soft,
and cannot provide significant evidence for variability,
 due to the small statistics and time resolution being
insufficient to look for the spin periods of the two pulsars.

A longer public observation (50 ks), carried out in 2004 March through the 
\xmm\ Director's Discretionary Time program, yielded an improvement in 
statistics by a factor $\sim$10 with respect to the first look of \cha. These 
data confirmed the softness of the spectrum, which could be fit either by a 
power law with photon index $\Gamma=3.5^{+0.5}_{-0.3}$ and absorption
$N_{\rm{H}}\simeq7.0\times10^{20}$ cm$^{-2}$ or by a blackbody with 
temperature of 0.15 keV and a lower interstellar absorption 
\citep{pellizzoni04}. These authors could also have performed the first X-ray timing 
analysis for \psr, but no periodic or aperiodic variations were found, with 
upper limits of $\sim$60\% on the pulsed fractions of both pulsars and of 
$\sim$40\% for modulations in the orbital period (all limits are at the 99\% confidence level 
and for sinusoidal light curves). \citet{campana04}  reported a joint spectral 
analysis of the \xmm\ (only the MOS) and  \cha\ data, reaching similar 
conclusions on the source spectrum.

A further \cha\ observation ($\sim$90 ks) with the  High-Resolution Camera showed X-ray 
pulses at the period of \psra\ \citep{chatterjee07}, with a double-peaked 
profile, similar to that observed in radio, and a pulsed fraction of 
$\sim$70\%. Although purely non thermal emission is consistent with the data, 
the X-ray pulse morphology of \psra, in combination with previously reported
spectral properties of the X-ray emission, suggests the existence of both 
non thermal magnetospheric emission and a broad sinusoidal thermal emission 
component from the neutron star surface. No pulsations were detected from 
pulsar B, nor evidence for orbital modulation.
Here we report the results of a $\sim$230 ks long X-ray observation of \psr\ 
obtained  within the frame of \xmm\ ``Large Programs'' in  2006 October.

\section{X-ray observation and Data Reduction}\label{reduction}

Our observation of \psr\ was carried out in two consecutive \xmm\ orbits. The 
first part of the observation started on 2006, October 26 at 
00:28:15 UT and lasted 119.8 ks, the second one started on October 
28 at 00:28:37 UT and lasted 114.5 ks. In total, the observation 
allowed the coverage of $\sim$26 revolutions of the binary system 
($P_{\rm{orb}}=8,834.535$ s). The pn camera \citep{struder01} was operated in 
Small Window mode (imaging across a $4'\times4'$ field of view with a 5.67 ms 
time resolution) with the medium optical filter. The MOS cameras 
\citep{turner01} were also set in the Small Window mode (yielding a 
$2'\times2'$ field of view in the central CCD, with a time resolution of 0.3 s)
with the medium filter. The data were processed using standard pipeline tasks 
(emproc and epproc) of the \xmm\ Science Analysis 
Software version 7.1.0.

In view of the faintness of \psr\ in soft X-rays, particular care in 
selecting source photons and reducing background contamination is crucial. We 
used only photons with pattern 0--12 for the MOS cameras, while for the pn we 
used pattern 0--4 for energies above 0.4 keV and pattern 0 for the $E<0.4$ keV 
energy range. Such an event selection allows one to reduce  by a factor $\sim$3.5 
the background count rate in the 0.15--0.4 keV energy range in the pn camera, 
while leaving almost unchanged the source count rate.

We evaluated the optimal selection of source events by maximizing the source 
signal-to-noise ratio in the 0.15--10 keV range, as a function of:

\emph{Source extraction region.}---We considered different extraction radii in the
10$\arcsec$--30$\arcsec$ range; the background was extracted in each case from 
two rectangular regions located at the same distance from the readout node as 
the source region.

\emph{Threshold for high particle background screening.}---The observations are 
affected by a few short soft proton flares. We extracted a light curve in the 
0.15--10 keV range with a 10 s time bin from the whole field of view for each 
camera. Following the prescription by \citet{deluca04}, we identified the 
quiescent average count rate and considered different thresholds in the
2--$8\,\sigma$ range from this average rate.\\

The optimal choice turned out to be a source extraction radius of $18\arcsec$ 
for the pn and the MOS\,1 cameras and of $15\arcsec$ for the MOS\,2 camera 
(which has a higher low-energy background) and a threshold at $5\,\sigma$ from
the quiescent rate to screen from high particle background episodes. The 
resulting source and background count statistics are shown in 
Table~\ref{counts}.

\begin{deluxetable}{lcccc}
\tabletypesize{\scriptsize}
\tablecolumns{1}
\tablewidth{0pt}
\tablecaption{\label{counts} Photon Harvest from the \xmm\ Observations of \psr.}
\tablehead{
\colhead{} & \colhead{} & \colhead{Total } & \colhead{Background} & \colhead{Exposure}\\
\colhead{Orbit} & \colhead{Detector} & \colhead{Counts\tablenotemark{a}} & 
\colhead{\phantom{1}fraction\tablenotemark{b}(\%)} & \colhead{\phantom{1}time\tablenotemark{c} (ks)}
}
\startdata
1260.......... & pn & 2483 & 24.5 & \phantom{1}80.7 \\
1260.......... & MOS\,1 & \phantom{1}652 & 21.7 & 111.5 \\
1260.......... & MOS\,2 & \phantom{1}750 & 26.1 & 112.3 \\
1261.......... & pn & 2222 & 23.6 & \phantom{1}74.9 \\
1261.......... & MOS\,1 & \phantom{1}587 & 19.7 & 102.9 \\
1261.......... & MOS\,2 & \phantom{1}697 & 29.5 & \phantom{1}\phantom{1}\,\,103.9\smallskip
\enddata
\tablenotetext{}{\textsc{Note.}---See text for details on data reduction and background screening.}
\tablenotetext{ a}{ Total number of counts inside the source extraction region.}
\tablenotetext{ b}{ Contribution of background to total number of counts.}
\tablenotetext{ c}{ Good observing time after dead-time correction and screening for soft-proton flares.}
\end{deluxetable}

\section{Timing Analysis}\label{timing}

\subsection{Pulsations from \psr A}\label{timinga}

For the search of pulsations at the spin period of pulsar A we could only use 
the pn data ($\sim$4700 counts), owing to the inadequate time resolution of 
the MOS data.
The times of arrival were converted to the solar system barycenter and 
corrected for the orbital motion and relativistic effects of the binary system 
according to \citet{blandford76}. This was done with a program we wrote  
ad hoc and tested using an \xmm\ observation of the binary millisecond pulsar 
XTE\,J1751$-$305 \citep{miller03}. As a further cross-check, we also compared 
the rotational phases obtained by our program with those resulting from the 
Tempo timing analysis software.\footnote{See \url{http://www.atnf.csiro.au/research/pulsar/tempo}.} The maximum discrepancy 
between timing corrections from our program and Tempo is at most  a 
few microseconds.

In our search for pulsations we  employed the $Z_n^2$ test \citep{buccheri83}, 
where we indicate with $n$ the number of harmonics. We examined a wide 
frequency range centered on the value predicted at the epoch (MJD 54\,034) of our \xmm\ 
observation by the radio measurements of \citet{kramer06}. The most significant
$Z_n^2$ statistics occurred for $n=1$ at $P_{\rm{A}}^{\rm{best}} =
22.6993787(5)$ ms (1$\,\sigma$ errors in the last digit are quoted in 
parentheses). The corresponding $Z_1^2$ value  is 378.91, which even when taking
into account the $\sim$$10^4$ searched periods,  has a virtually null
probability of chance occurrence. Our best-fit period is consistent with the 
value $P_{\rm{A}}^{\rm{radio}} = 22.699378466112(5)$ ms expected from the radio
ephemeris of \citet{kramer06}.

\begin{figure}
\resizebox{\hsize}{!}{\includegraphics[angle=00]{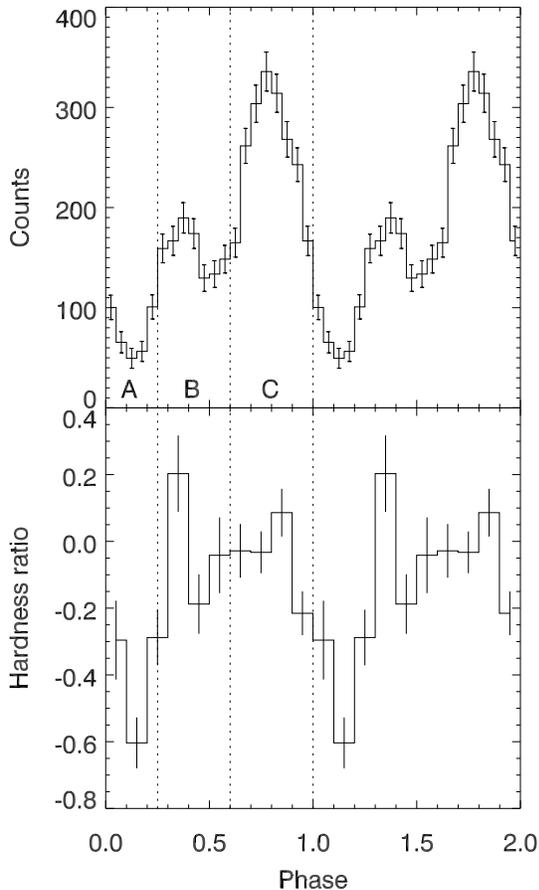}}
\caption{\label{fig1} Background-subtracted pn light curve of \psra\ in the 0.15--4 keV energy
range folded at period $P_{\rm{A}}^{\rm{best}} =22.6993787(5)$ ms (\emph{top}) and corresponding hardness ratio curve based on the soft (0.15--0.3
keV) and hard (0.8--3 keV) energy ranges (\emph{bottom}). The dashed lines
indicate the three phase intervals (A, B, and C) used for the phase-resolved
spectroscopy.}
\end{figure}

The background-subtracted and exposure-corrected light curve of pulsar A in the
0.15--4 keV energy range is shown in the top panel of Figure~\ref{fig1} (due to 
the soft pulsar spectrum, the signal-to-noise ratio above 4 keV is very small). 
Error bars are calculated according to the expression
$\sigma_{i}=(C_{i}+B_{i} f^{2})^{1/2}/E_{i}$, where $C_{i}$, $B_{i}$ and 
$E_{i}$ are respectively the total counts, background counts, and exposure in 
the bin $i$, and $f$ is the ratio between the source and background extraction area
($f\simeq0.11$). We checked that the pulsations are significantly detected down
to the lowest energy bins  covered by the pn instrument; in fact, the Pearson 
statistics for a 10 bin light curve in the 0.15--0.2 keV range gives a reduced
$\chi^2$ (hereafter $\chi^2_r$) of 3.85, corresponding to a 4$\,\sigma$ 
detection.

The pulse profile is double peaked and deeply modulated, with a pulsed fraction
of $75.7\%\pm5.4\%$ in the energy range 0.15--4 keV. To calculate the pulsed 
flux, we considered all the counts above the minimum of the light curve, using 
the expression $PF\equiv C_{\rm{tot}}-n N_{\rm{min}}$ and its associated error
$\sigma_{PF}=(C_{\rm{tot}}+n^2 \sigma^2_{N_{\rm{min}}})^{1/2} \simeq n 
(N_{\rm{min}})^{1/2}$, where $C_{\rm{tot}}$ are the total counts, $n$ is the  
number of bins in the light curve and $N_{\rm{min}}$ are the counts of the 
minimum. The expression $(C_{\rm{max}}-C_{\rm{min}})/(C_{\rm{max}}+
C_{\rm{min}})$ used by \citet{chatterjee07} does not account for the pulsed 
flux associated with secondary peaks; nevertheless their pulsed fraction value is
in agreement with our result within 1$\,\sigma$. Both methods are ``bin 
dependent,'' but reasonable different choices of the number of bins (i.e. 
$n>10$) do not significantly affect the results. Note that the pulsed fraction 
upper limit of 60\% obtained in the short 2004 \xmm\ observation referred to a
sinusoidal profile \citep{pellizzoni04}, and therefore it is not inconsistent 
with the present result.

\begin{figure*}
\centering
\resizebox{.8\hsize}{!}{\includegraphics[angle=00]{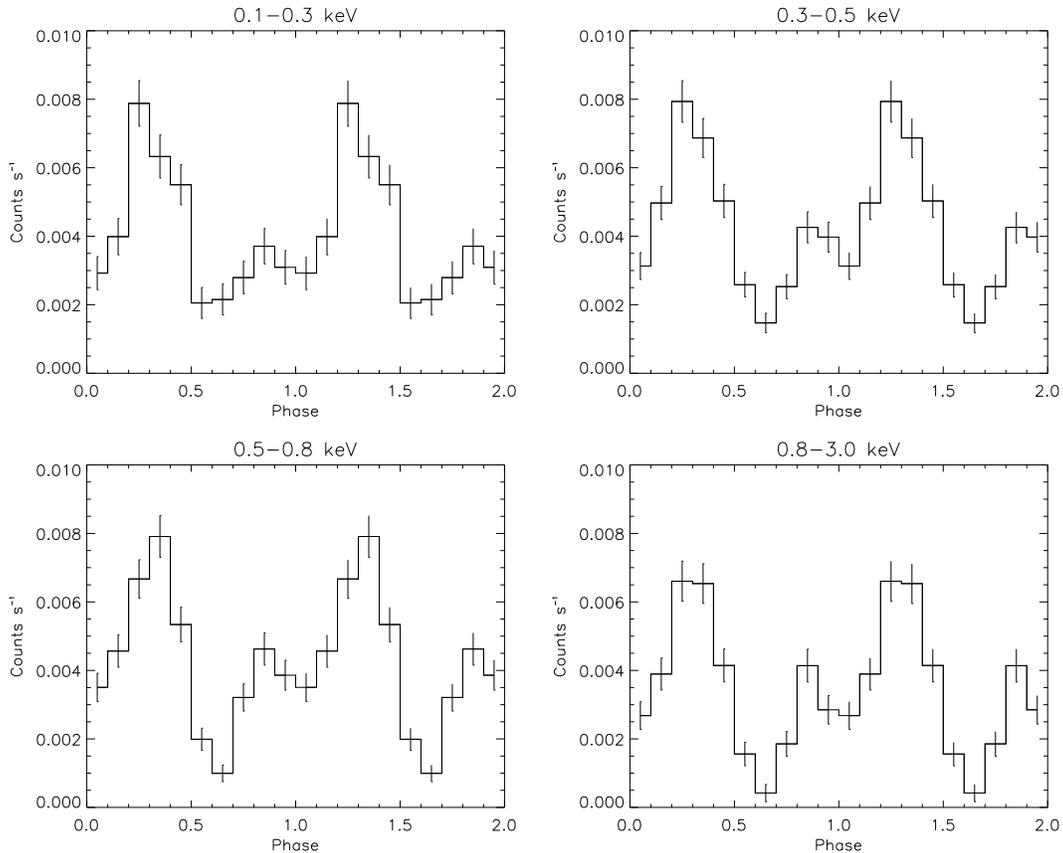}}
\caption{\label{fig2} Background-subtracted light curves of \psra\ for four 
energy intervals providing 800--900 counts each. The count rate 
of the minimum of the light curve is compatible with 0 at high energies, while at 
low energies the presence of significant unpulsed flux is apparent.}
\end{figure*}

\begin{figure}
\resizebox{\hsize}{!}{\includegraphics[angle=00]{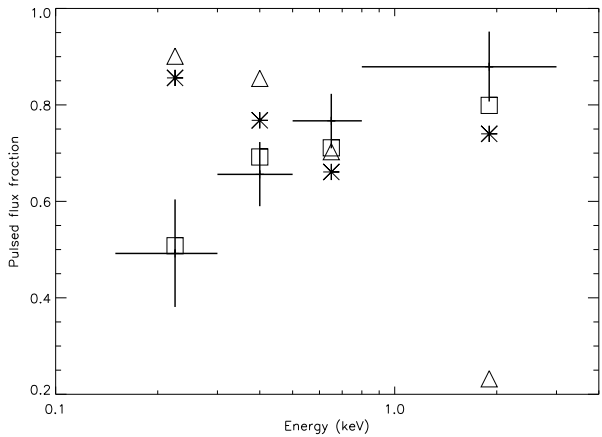}}
\caption{\label{fig3} Pulsed fraction of \psra\ (\emph{crosses}) as a function of 
energy (energy intervals as in Figure~\ref{fig2}). The plot also shows the 
predicted pulsed flux (\emph{triangles, stars, and squares}) for the spectral models 
discussed in \S\S ~\ref{spec} and \ref{discussion}.}
\end{figure}

A simple hardness ratio analysis, based on the soft ($S$: 0.15--0.3 keV) and 
hard ($H$: 0.8--3 keV) energy ranges,\footnote{The hardness ratio is defined as 
$(H-S)/(H+S)$, where for each phase bin $H$ and $S$ are the background-subtracted count rates in the hard and soft ranges.} indicates a softer 
spectrum in correspondence to the minimum in the folded light curve (see bottom
panel of Figure~\ref{fig1}). To investigate the energy dependence of the pulsed 
fraction in more detail and as a cross-check of complementary phase-resolved 
spectral studies discussed below, we produced the folded light curves for four 
different energy ranges (0.15--0.3 keV, 0.3--0.5 keV, 0.5--0.8 keV, and 0.8--3 keV)
providing 800--900 source counts each (Figure~\ref{fig2}). The corresponding 
pulsed fractions, plotted in Figure~\ref{fig3}, increase from 50\% to 90\% from 
lower to higher energies. These values have a probability smaller than 1\% of 
deriving from an energy-independent distribution.

\begin{figure}
\resizebox{\hsize}{!}{\includegraphics[angle=00]{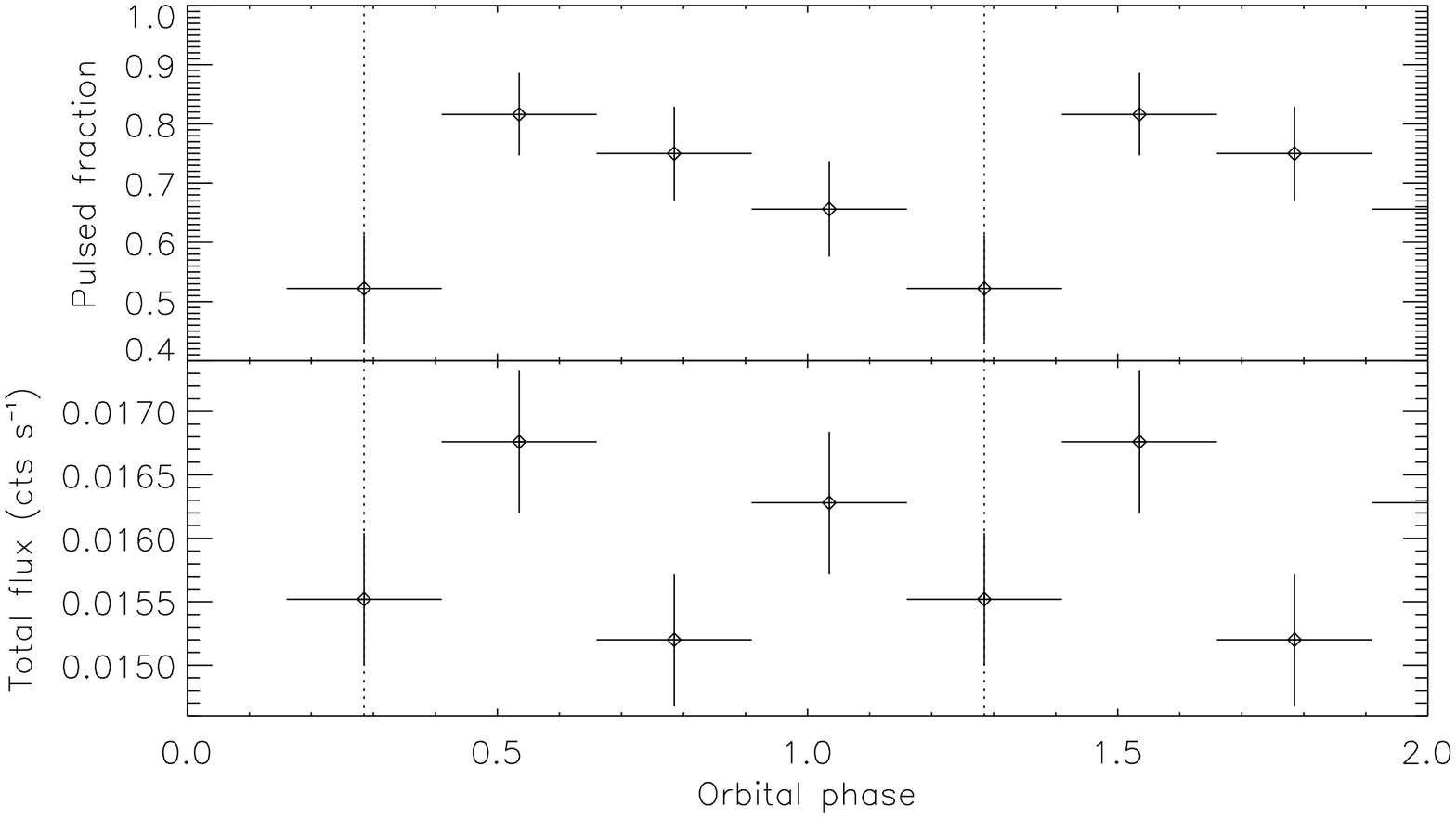}}
\resizebox{\hsize}{!}{\includegraphics[angle=00]{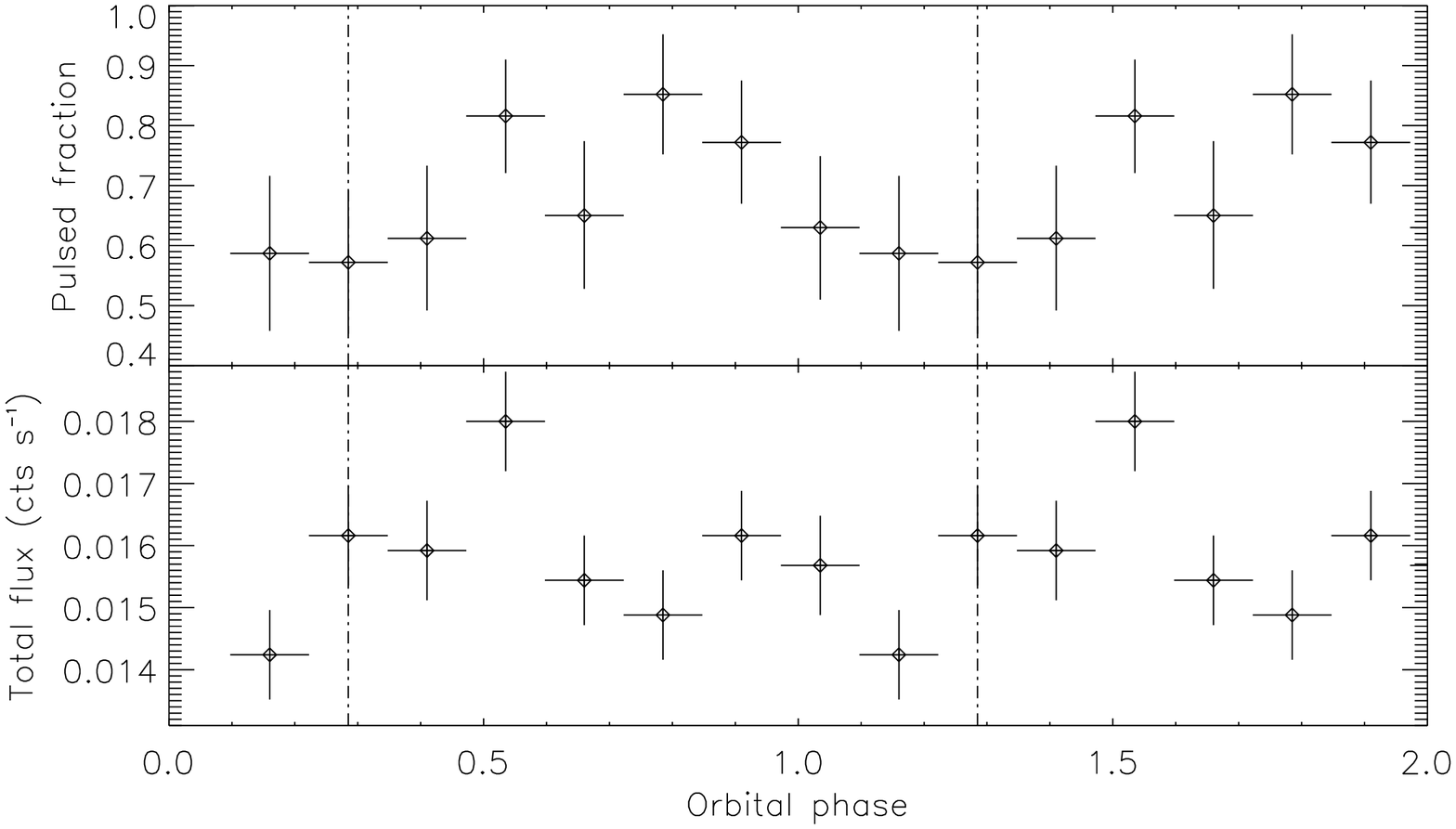}}
\caption{\label{fig5} Pulsed fraction and total flux of \psra\ as a function of orbital phase integrated on $\sim$40 min (\emph{top})
and $\sim$20 min (\emph{bottom}) bins. The dotted vertical line corresponds
to the superior conjunction of A (when A is occulted by B).}
\end{figure}

We searched for possible flux and pulse profile variations due to the pulsars' 
mutual interaction as a function of the orbital phase. For example, if there is
a bow shock structure as that described in \citet{lyutikov05}, associated with an
unpulsed variable component (see \S ~\ref{discussion}), we would expect 
variations in the total flux correlated with changes in the pulsed fraction. In
Figure~\ref{fig5} (\emph{top}) we report the pulsed fraction and total flux obtained from the
folded 0.15--3 keV light curves integrated in four orbital phase intervals. The
pulsed fraction varies in the 50\%--80\% range with the minimum corresponding to
the superior conjunction of A (when A is occulted by B; \emph{dotted line}). However,
the values are consistent at the 5.5\% level with a uniform distribution and 
there is no corresponding change in the total flux. A similar analysis dividing
the orbit in eight parts (Figure~\ref{fig5}, \emph{bottom}) gave only marginal evidence 
of a $\sim$15\% variability (null hypothesis probability null hypothesis probability = 1.5\%) in 
the pn flux.

To calculate upper limits on orbital flux modulation, we considered the 
fraction of the counts above the minimum of the light curve with error 
evaluation similar to those described in this subsection. For time scales
of $\sim$15 min (10 bins) the 1$\,\sigma$ upper limits on variability (pn+MOS data) is 
of 11.5\% for the 0.15--4 keV range and $<$20\%--30\% for the selected bands
mentioned above. Longer time-scales of 0.5--1 hr imply upper limits 
$\leqslant$10$\%$ on all energy selections.
The search for orbital and aperiodic variability, even when selecting
the time intervals corresponding to the minimum in the \psra\ folded light 
curve, does not improve the above upper limits.

\subsection{Pulsations from \psr B}\label{timingb}

We searched for pulsations from \psrb\ using the same method and radio 
ephemeris reference as for \psra\ \citep{kramer06}, but in this case, owing to 
the longer pulse period, we could also use the lower time resolution MOS data.
The expected pulsar period at the epoch of our observation is
$P_{\rm{B}}^{\rm{radio}}=2.7734607024(7)$ s, with an uncertainty orders of 
magnitude smaller than the intrinsic pulse search resolution of our data set, 
$\frac{1}{2}P^2_{\rm{B}}/T_{\rm{OBS}} \simeq 1.4 \times 10^{-5}$ s. Thus, if 
the rotational parameters of the neutron star are stable (as suggested by radio ephemeris), a single-trial pulse 
search at $P_{\rm{B}}$ is appropriate. However, since we cannot in principle exclude that (unlikely) 
glitches and/or significant timing noise occurred in the time between the radio
and X-rays observations, we also performed a search in a range of periods
around $P_{\rm{B}}^{\rm{radio}}$. In any case, no significant pulsations were
detected in the whole energy range nor in our canonical bands (0.15--0.3 keV, 
0.3--0.5 keV, 0.5--0.8 keV, and 0.8--3 keV) providing 1200--1400 source counts 
each. The upper limit on the pulsed flux fraction is of 40\%.

Since the radio flux of \psrb\ is strongly modulated as a function of the 
orbital phase and is nearly disappearing for most of the orbit 
\citep{lyne04,burgay05}, we also performed an orbital phase-dependent pulse 
search.
We folded the counts at $P_{\rm{B}}^{\rm{radio}}$ for four orbital phase 
intervals around the pulsars' conjunctions ($\Delta\phi=[$0.16--0.41] and 
[0.66--0.91]) and quadratures ($\Delta\phi=[$0.41--0.66] and 
[0.91--1.16]).  As a reference, consider that orbital phase 0.28 corresponds to the superior conjunction of A. A positive detection of the \psrb\ pulsations in 
the pn data set is obtained for the phase interval corresponding to the 
ascending quadrature of B ($\Delta\phi=[$0.41--0.66]). The Pearson statistics for the pulse histogram with
20 bins ($\sim$250 mean source counts per bin) in the 0.15--3 keV band provides
a $\chi^2_r$ of 2.4 (19  degrees of freedom [dof]) corresponding to
a probability of only 0.06\% that the profile is drawn from a uniform 
distribution ($3.4\,\sigma$). A further analysis with the $Z^2_n$ test supports
the detection; with the number of harmonics $n$ being varied from 1 to 5, we 
found a statistically significant signal for $n=4$ ($Z^2_4=34.55$, 
4.2$\,\sigma$) and 5 ($Z^2_5=35.48$, 3.8$\,\sigma$). No significant 
detections are obtained for the other orbital phase intervals ($\chi_r^2<1.4$, 
19 dof).

We tried to better constrain the orbital phase interval corresponding to a 
positive detection of \psrb\  by shifting and/or shortening the orbital phase 
intervals of the period search. However we did not find more significant 
detections, and considering the low-statistics pulsed data, we cannot derive
tighter constraints.

\begin{figure}
\resizebox{\hsize}{!}{\includegraphics[angle=00]{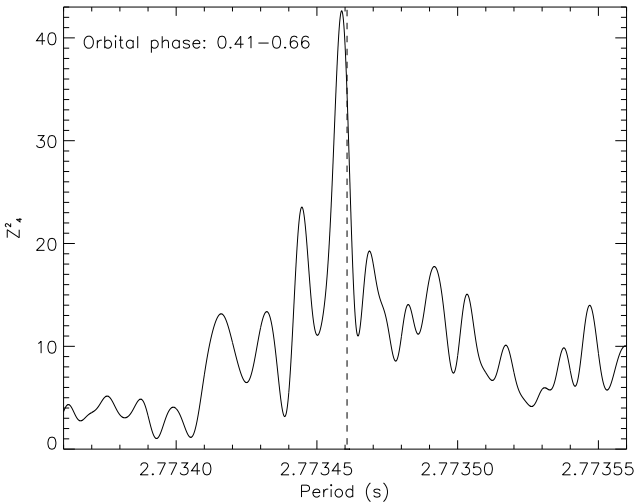}}
\caption{\label{psrB} Distribution of the $Z_4^2$ statistics for the pn data 
of \psr\ during the orbital
phase interval $\Delta\phi=[0.41$--$0.66]$ (see \S ~\ref{timingb} for
details). The vertical dashed line indicates the \psrb\ period expected from
adopting the radio ephemeris of \citet{kramer06}.}
\end{figure}

As mentioned above,  the stability of the timing  parameters of this neutron 
star is likely, but not guaranteed. Thus, in principle, a strict single-trial search at the
predicted value $P_{\rm{B}}^{\rm{radio}}$ might not be the optimal strategy.
Furthermore, a blind search in period is an important cross-check to claim a robust pulsed detection. 
We therefore also searched for other peaks in the $Z^2_n$  distributions around 
$P_{\rm{B}}^{\rm{radio}}$. We did not find more significant detections apart 
from values close (within the pulse search resolution) to 
$P_{\rm{B}}^{\rm{radio}}$ (Figure~\ref{psrB}). The best-fit pulse period is 
$P_{\rm{B}}^{\rm{best}} =2.773459(4)$ s providing a probability of only 
$\sim$$10^{-6}$ ($Z_4^2=42.63$, 4.9$\,\sigma$) that we are sampling a uniform 
distribution.

\begin{figure*}
\centering
\resizebox{.8\hsize}{!}{\includegraphics[angle=00,width=15cm]{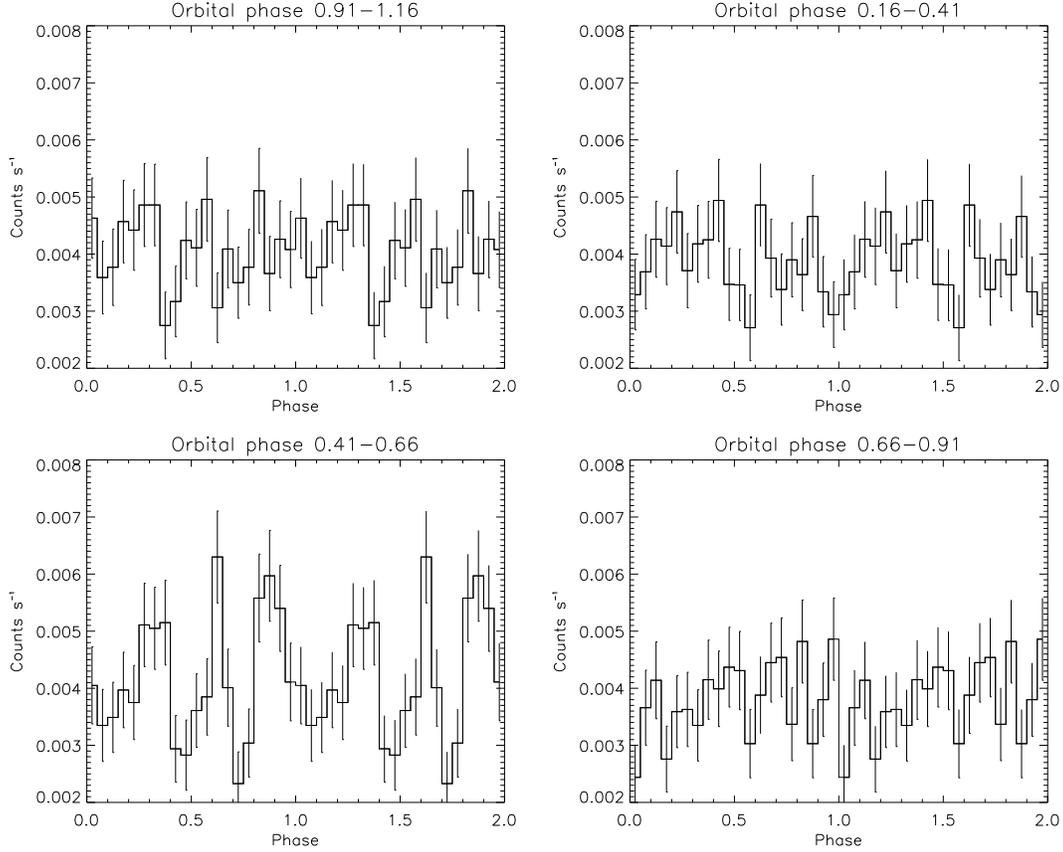}}
\caption{\label{timingB2} Light curves of \psrb\ folded at period
$P_{\rm{B}}^{\rm{best}} =2.773459(4)$ s
as a function of the orbital phase. The pulsar is detected at the 4.9$\,\sigma$ level
around the ascending node (orbital phase interval 0.41--0.66).}
\end{figure*}

The resulting pn light curves for the four phase intervals considered are shown
in Figure~\ref{timingB2}. The pulse profile appears highly structured and with a
pulsed flux fraction of $35\pm15$\% in the energy range 0.15--3 keV. 

The detection of  \psrb\ in MOS data is only marginal, and the analysis of 
merged MOS and pn data does not provide better detection significance (even when
selecting counts which are in the lowest emission bins of the pulse profile of 
\psra), probably because the 0.3 s time resolution of the MOS dilutes the sharp
peaks of the light curve. Furthermore, the poor effective area of the MOS at 
low energies (i.e. negligible in the 0.15--0.3 keV band) is not efficient for 
pulse searches in sources with very soft spectra. In fact, although \psrb\ is 
also marginally detected at $E>0.5$ keV ($\chi_r^2=2.1$, 19 dof), most of the
flux contributions to the main peaks of its light curve come from the low-energy band.

In the hypothesis that \psrb\ is emitting X-rays only in the orbital phases in 
which we can firmly claim pulse detection, an orbital flux variability with 
magnitude comparable with the pulsed flux fraction should be present. In fact,
there is marginal evidence that the pn total flux peaks around $\phi\sim0.5$ 
when B's pulses appear (Figure~\ref{fig5}). It is worth noting that the sum of 
the pulsed emission of the two pulsars roughly accounts for the whole X-ray 
flux, not allowing for a significant amount of emission from other sources (e.g. 
shock emission from the pulsars' mutual interactions).

\section{Spectral analysis}\label{spec}

Source and background spectra were extracted from the pn and MOS data using the
regions and soft protons filtering criteria described in  \S ~\ref{reduction}. For what concerns the pn pattern selection, we used only 
single events (pattern 0) for the whole energy range in order to reduce the 
background in the lowest energy channels. We checked that fully consistent 
results were obtained including double events (pattern 0--4).

Each spectrum was fitted with the XSPEC version 11.3.2 software 
package using four models, all modified by photoelectric absorption 
(phabs model with cross section from \citealt{balucinska92} and 
abundances from \citealt{anders89}): power law (PL), blackbody (BB), two 
blackbodies (BB$_{\rm{C}}$+BB$_{\rm{H}}$), and power law plus blackbody 
(PL+BB). The best-fit parameters for the pn spectrum  and for a simultaneous 
fit of the pn, MOS\,1, and MOS\,2 spectra are reported in Table~\ref{fits}. 
\begin{deluxetable*}{lccccccccc}
 \tabletypesize{\scriptsize}
  \tablecolumns{2}
\tablewidth{0pc}
 \tablecaption{Summary of the Spectral Results in the
0.15--10 $\rm{keV}$ Energy Range.\label{fits}} \tablehead{
\colhead{} & \colhead{} & \colhead{$N_{\rm H}$} & \colhead{$\Gamma$} & \colhead{} & \colhead{$k_B T_{\rm{BB_H}}$} & \colhead{$R_{\rm{BB}_{\rm{H}}}$\tablenotemark{b}} & \colhead{$k_B T_{\rm{BB_C}}$} & \colhead{$R_{\rm{BB}_{\rm{C}}}$\tablenotemark{b}} & \colhead{$\chi^{2}_{r}$ (dof)}\\
\colhead{Instrument} & \colhead{Model} & \colhead{($10^{20}\rm cm^{-2}$)} &
\colhead{} & \colhead{PL norm.\tablenotemark{a}} & \colhead{(eV)} & \colhead{(m)} &
\colhead{(eV)} & \colhead{(km)} & \colhead{} } \startdata
pn......................... & PL  & 4.6$^{+0.8}_{-0.7}$ & $3.4\pm0.1$ & 9.4$\pm$0.6 & \nodata & \nodata & \nodata & \nodata & 1.38 (50) \\
 & BB  & 0 & \nodata & \nodata & 142 & 149 & \nodata & \nodata & 3.29 (50) \\
 & BB$_{\rm{H}}$+BB$_{\rm{C}}$  & $<0.9$ & \nodata & \nodata & 280$^{+50}_{-40}$ & 23$^{+12}_{-7}$ & 110$^{+8}_{-13}$ & 0.23$^{+0.05}_{-0.03}$ & 1.14 (48) \\
 & PL+BB  & $3.2^{+1.3}_{-1.4}$ & $3.3^{+0.3}_{-0.4}$ & 6.6$^{+1.4}_{-1.3}$ & 160$^{+40}_{-20}$ & 60$^{+50}_{-30}$ & \nodata & \nodata & 1.20 (48) \\
 pn+MOS............. & PL  & 5.0$^{+0.7}_{-0.6}$ & 3.4$\pm0.1$ & 9.6$\pm0.4$ & \nodata & \nodata & \nodata & \nodata & 1.16 (98) \\
 & BB  & 0 & \nodata & \nodata & 147 & 138 & \nodata & \nodata & 3.05 (98) \\
 & BB$_{\rm{H}}$+BB$_{\rm{C}}$  & $<0.5$ & \nodata & \nodata & 290$^{+40}_{-30}$ & 20$^{+8}_{-6}$ & 114$^{+7}_{-9}$ & 0.21$^{+0.04}_{-0.02}$ & 1.00 (96) \\
 & PL+BB  & 3.0$\pm$1.1 & $3.2^{+0.1}_{-0.3}$ & 6.5$\pm$1.0 & $150\pm$20 & 80$\pm$30 & \nodata & \nodata & \phantom{1}\phantom{1}0.92 (96) \smallskip
\enddata
\tablenotetext{}{\textsc{Note.}---Errors are given at the 90\% confidence
level and are not reported for the single-blackbody model because
the fit is largely unacceptable.}
\tablenotetext{ a}{ In units of 10$^{-6}$ photons cm$^{-2}$ s$^{-1}$
keV$^{-1}$, at 1 keV.} 
\tablenotetext{ b}{ Assuming a distance of 500
pc.}
\end{deluxetable*}

In contrast with the previous short \cha\ and \xmm\ observations 
\citep{mclaughlin04,pellizzoni04,campana04}, single-component models fail to 
properly fit the data. A blackbody model is completely ruled out and only 
a very steep power law provides a barely acceptable fit (null hypothesis probability $= 4$\%). 
Either a double-blackbody or a power law plus blackbody model are instead an 
adequate fit (null hypothesis probability $>$ 50\%). In the two-component models, the derived 
column density is also in better agreement with the value estimated from the 
radio pulsars' dispersion measure ($N_{\rm{H}}=1.5\times10^{20}$ cm$^{-2}$, assuming 1 H for each electron; \citealt{campana04}). 
The power law photon index in the PL+BB model is still very soft. A hard 
power law with photon index $\Gamma\sim2$, as expected for a shock emission, 
does not provide an acceptable fit. The observed flux in the 0.2--3 keV energy 
range is $4.0\times10^{-14}$ \flux, corresponding to an unabsorbed luminosity 
L$_{\rm{X}}=2.2\times10^{30} d_{0.5}^2$ \lum\ for the PL+BB model. The radio 
dispersion measure distance of $\sim$500 pc is based on a model for the 
interstellar electron density \citep{cordes02}, and the parallactic distance 
ranges from 0.2 to 1 kpc \citep{kramer06}, implying uncertainties of a factor 
$\sim$4 in luminosity estimates.

The good timing resolution coupled to the simultaneous spectral capability of 
the pn data, together with the relatively large number of detected source 
photons, allowed us to perform for the first time a phase-resolved spectral 
analysis for \psra. Three spectra were extracted by selecting the pn events (with 
pattern equal to 0) in the three phase intervals shown in Figure~\ref{fig1}, 
corresponding to the off-pulse (phase A), the interpulse (phase B), and the 
main pulsation peak (phase C). The results of the fit of these three spectra 
with the four models used for the phase-averaged spectrum are shown in 
Table~\ref{fits-pr}.
\begin{deluxetable*}{lccccccccc}
 \tabletypesize{\scriptsize}
  \tablecolumns{2}
\tablewidth{0pc}
 \tablecaption{Results of the Phase-resolved Spectroscopy Performed with the pn Single Events in the
0.15--10 $\rm{keV}$ Energy Range. \label{fits-pr}} \tablehead{
\colhead{Model} & \colhead{Phase} & \colhead{$N_{\rm H}$} & \colhead{$\Gamma$} & \colhead{PL norm.\tablenotemark{a}} & \colhead{$k_B T_{\rm{BB_H}}$} & \colhead{$R_{\rm{BB}_{\rm{H}}}$\tablenotemark{b}} & \colhead{$k_B T_{\rm{BB_C}}$} & \colhead{$R_{\rm{BB}_{\rm{C}}}$\tablenotemark{b}} & \colhead{$\chi^{2}_{r}$ (dof)}\\
\colhead{} & \colhead{interval} & \colhead{($10^{20}\rm cm^{-2}$)} &
\colhead{} & \colhead{} & \colhead{(eV)} & \colhead{(m)} &
\colhead{(eV)} & \colhead{(km)} & \colhead{} } \startdata
PL..........................& A  & $2.1^{+2.2}_{-1.6}$ & $3.5^{+0.5}_{-0.3}$ & 2.9$^{+0.7}_{-0.6}$ & \nodata & \nodata & \nodata & \nodata & 1.37 (9) \\
  & B  & $4.9^{+1.6}_{-1.4}$ & 3.3$^{+0.3}_{-0.2}$ & 8.6$^{+1.0}_{-0.9}$ & \nodata & \nodata & \nodata & \nodata & 0.99 (23) \\
  & C  & $5.0^{+1.2}_{-1.1}$ & $3.4\pm0.2$ & 13.4$^{+1.5}_{-0.9}$ & \nodata & \nodata & \nodata & \nodata & 1.11 (42) \\
BB..........................& A  & 0 & \nodata & \nodata  & 118 & 149 & \nodata & \nodata &  2.57 (9) \\
  & B  & 0 & \nodata & \nodata  & 141 & 138 & \nodata & \nodata &  2.16 (23) \\
  & C  & 0 & \nodata & \nodata  & 146 & 167 & \nodata & \nodata &  2.67 (42) \\
BB$_{\rm{H}}$+BB$_{\rm{C}}$.............&A  & $9.0^{+3.2}_{-3.0}$ & \nodata & \nodata & $140\pm20$ & 130$\pm$40 & 29$^{+42}_{-2}$ & 35.8$^{+93.3}_{-35.4}$ & 0.94 (7) \\
 & B  & $<1.9$ & \nodata & \nodata & 370$^{+150}_{-100}$ & 11$^{+13}_{-5}$ & 120$^{+10}_{-20}$ & 0.18$^{+0.06}_{-0.02}$ & 0.86 (21) \\
 & C  & $<2.3$ & \nodata & \nodata & 260$\pm50$ & 34$^{+24}_{-12}$ & 100$^{+10}_{-20}$& 0.30$^{+0.31}_{-0.06}$ & 1.00 (40) \\
PL+BB...................& A  & $10\pm3$ & $9.9^{+14.3}_{-6.7}$  & 0.02$^{+0.70}_{-0.01}$ & $140\pm20$ & 120$^{+60}_{-40}$ & \nodata & \nodata & 1.14 (7) \\
 & B  & 2.0$^{+2.9}_{-1.8}$ & $2.7\pm0.7$ & 5.3$^{+2.8}_{-2.5}$ & 120$^{+30}_{-20}$ & 0.14$\pm$0.09 & \nodata & \nodata & 0.88 (21) \\
 & C  & $4.0\pm1.7$ & $3.5^{+0.6}_{-0.3}$ & 9.7$^{+2.8}_{-3.2}$ & 190$^{+20}_{-50}$ & 50$^{+50}_{-30}$ & \nodata & \nodata & \phantom{1}0.95 (40)\smallskip
\enddata
\tablenotetext{}{\textsc{Note.}---Errors are given at the 90\% confidence
level and are not reported for the single-blackbody model because
the fit is largely unacceptable.}
\tablenotetext{ a}{ In units of 10$^{-6}$ photons cm$^{-2}$ s$^{-1}$
keV$^{-1}$, at 1 keV.} 
\tablenotetext{ b}{ Assuming a distance of 500
pc.}
\end{deluxetable*}
\indent A single-blackbody model does not provide an acceptable fit for any of the 
three spectra. We also note that, while the best-fit parameters of the main 
peak (phase C) and interpulse (phase B) are very similar to those of the 
phase-averaged spectrum, the fits of the off-pulse spectrum (phase A) with the 
two-component models converge to rather different values for some parameters.
In particular, in the double-blackbody model, one of the two blackbodies has a 
temperature $\sim$10 times smaller (and a correspondingly larger emitting area)
than in the other two phase intervals. Moreover, in the PL+BB model the 
power law component is steeper, and in both cases, the column density is 
significantly larger.

\begin{figure}
\resizebox{\hsize}{!}{\includegraphics[angle=-90]{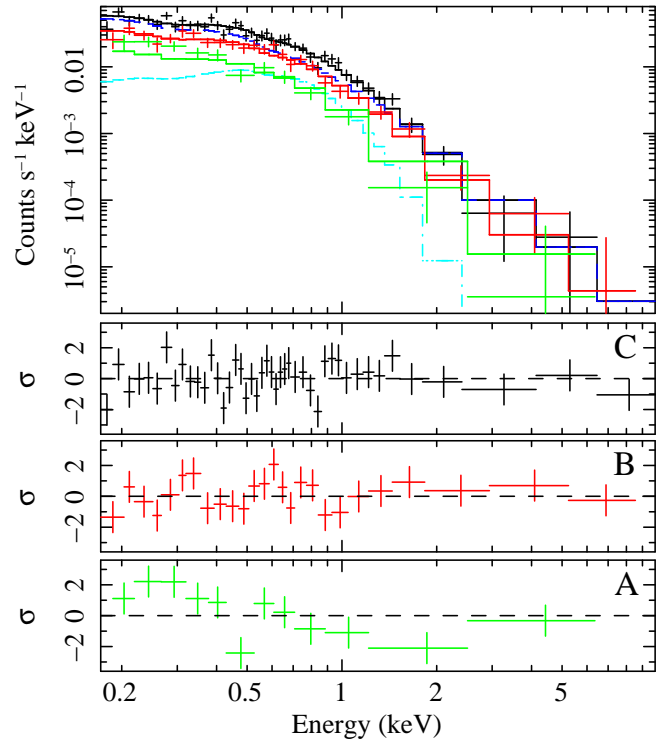}}
\caption{\label{pps_fig} Simultaneous fit of
the three pn phase-resolved spectra (see Figure~\ref{fig1} for the
corresponding phase intervals) with a power law plus blackbody
spectrum with linked parameters, except for an overall normalization
factor (see Table \ref{fits-pr}). For the spectrum of the pulse peak (\emph{black}), the relative contribution of the power law (\emph{blue}) and
blackbody (\emph{light blue}) components are also shown in the spectrum.
The residuals are shown separately for each spectrum in standard
deviation units.}
\end{figure}

To better check if the source spectrum is actually changing with the phase of 
\psra, we simultaneously fitted the three phase-resolved spectra with the same
blackbody plus power law model with all the parameters linked together, except 
for an overall normalization factor. A globally acceptable fit is obtained, but
if we compute separately the $\chi^{2}_{r}$ for each, fixing all the parameters
to the best-fit values of the simultaneous fit, we note that the fit of the 
off-pulse spectrum is unacceptable (see Table~\ref{tab_new}). The corresponding
residuals (see Figure~\ref{pps_fig}) show a soft excess in the off-pulse 
spectrum, confirming the softening in the pulse minimum already displayed in 
the hardness ratio (see Figure~\ref{fig1}).

\begin{deluxetable*}{cccccccccccc}
 \tabletypesize{\scriptsize}
  \tablecolumns{2}
\tablewidth{0pc}
 \tablecaption{Results of Simultaneous Fit of the Three Phase-resolved pn Spectra of \psra.\label{tab_new}} \tablehead{
\colhead{Fixed} & \colhead{Variable} & \colhead{Phase} & \colhead{$N_{\rm H}$} & \colhead{} & \colhead{PL} & \colhead{$k_B T_{\rm{BB_H}}$} & \colhead{$R_{\rm{BB}_{\rm{H}}}$\tablenotemark{b}} & \colhead{$k_B T_{\rm{BB_C}}$} & \colhead{$R_{\rm{BB}_{\rm{C}}}$\tablenotemark{b}} & \colhead{$\chi^{2}_{r, \rm single}$}& \colhead{$\chi^{2}_{r, \rm tot}$}\\
\colhead{component} & \colhead{component}  & \colhead{interval}&
\colhead{($10^{20}\rm cm^{-2}$)} & \colhead{$\Gamma$} & \colhead{\phantom{1}norm.\tablenotemark{a}} &
\colhead{(eV)} & \colhead{(m)} & \colhead{(eV)} & \colhead{(km)} &
\colhead{(dof)} & \colhead{(dof)}} \startdata
&  & A & &  & 3.0$^{+0.9}_{-0.8}$ & & 45$^{+30}_{-20}$ & \nodata & \nodata & 2.15 (12)& \\
\nodata  &  PL+BB\tablenotemark{c}  & B & $3.2^{+1.4}_{-1.3}$ & $3.2^{+0.2}_{-0.3}$ & 5.9$\pm$1.5 & 150$^{+50}_{-30}$ & 60$^{+40}_{-30}$ & \nodata & \nodata &  0.84 (26) & 1.15 (76)\\
 &   & C &  &  & 9.8$^{+1.6}_{-1.9}$ & & 80$^{+50}_{-30}$ & \nodata & \nodata &  0.89 (45) &\\
 & & A &  & & $<$2.2 & & & & & 0.71 (12) &\\
BB$_{\rm{H}}$+BB$_{\rm{C}}$ & PL & B & 6.9$^{+1.5}_{-1.1}$ & 3.3$^{+0.1}_{-0.2}$& 6.0$^{+1.3}_{-1.4}$ & $134^{+17}_{-14}$& 110$\pm40$ & 32$^{+5}_{-4}$& 15$^{+4}_{-5}$& 0.96 (26) & 1.00 (74)\\
 &  & C & & & 12.8$^{+0.8}_{-1.6}$ & & & & & 0.91 (45) &\\
& & A & & & & & $<$45 & & & 1.70 (12) &\\
PL+BB$_{\rm{C}}$& BB$_{\rm{H}}$& B &$<$1.6 & 2.2$^{+0.5}_{-0.7}$ &2.3$^{+1.0}_{-1.3}$ &166$\pm$14 & 70$^{+20}_{-10}$ & 84$^{+14}_{-24}$&0.26$^{+0.4}_{-0.05}$ & 0.94 (26)& 1.37 (74)\\
& & C & & & & & 110$^{+30}_{-20}$ & & & 1.25 (45) & \smallskip
\enddata
\tablenotetext{}{\textsc{Note.}--- Errors are given at the 90\% confidence
level.}
\tablenotetext{ a}{ In units of 10$^{-6}$ photons cm$^{-2}$ s$^{-1}$
keV$^{-1}$, at 1 keV.} 
\tablenotetext{ b}{ Assuming a distance of 500 pc.} 
\tablenotetext{ c}{ The relative normalization between the two
components is fixed.}
\end{deluxetable*}

We also tested the possibility that separate physical processes, producing 
different spectral components, were responsible for the pulsed and unpulsed
emission. A first hypothesis, already proposed by \citet{chatterjee07}, where a
blackbody accounts for the unpulsed emission and a power law for the pulsation 
is immediately discarded by the fact that the off-pulse spectrum cannot be well
fitted by a single-blackbody model (see Table~\ref{fits-pr}).

The off-pulse spectrum is instead well fitted by two-blackbodies. Therefore we 
tried to fit the phase-resolved spectra with a three-component model composed 
of two constant blackbodies and a power law with stable photon index and 
variable normalization. The resulting fit is very good and its parameters are
shown in Table~\ref{tab_new}.

Since a good fit of the off-pulse spectrum is also obtained using a PL+BB model
(see Table~\ref{fits-pr}), we also tested a model where the unpulsed spectrum 
was composed by a blackbody plus a power law and a second blackbody was 
responsible for the pulsed emission. In this model, the two blackbodies might 
come from the surface of the two pulsars and the power law from a shock. 
However, the resulting fit, also reported in Table~\ref{tab_new}, is not 
acceptable.

\section{Discussion}\label{discussion}

Our deep \xmm\ observation of the double-pulsar \psr\ reveals a complex X-ray 
phenomenology that cannot be simply ascribed to \psra\ alone, although it is 
clear that the only significant power plant of the system (apart from its 
strong gravitational potential) is the spin-down energy release of this pulsar.

We confirmed deeply modulated pulsed X-ray emission from \psra\ and we also 
detected for the first time X-ray pulsations from \psrb. The rotational energy 
loss of \psrb\ is by far insufficient to produce the observed X-ray luminosity.
It can also be excluded that the observed \psrb's X-rays derive from residual 
internal heat of this 50 million years old pulsar. Therefore, this detection 
gives  strong evidence for the mutual interactions of the two pulsars also at 
high energies. Although, as discussed below, there is no direct evidence of a 
termination shock between their winds and magnetospheres, we interpret the 
\psrb\ emission, visible only at the orbital phases in which the \psra\ beam 
intercepts \psrb, as ultimately powered by \psra.

We found that double-component models are required to account for the spectral 
phenomenology of the double-pulsar X-ray emission. A double-blackbody 
(BB$_{\rm{C}}$, $k_BT=114$ eV and BB$_{\rm{H}}$, $k_BT=290$ eV) provides a good
fit, while a  PL+BB model with a blackbody temperature of $k_BT=150$ eV and a 
soft photon index of 3.2 is even better.

It is interesting to compare the relative contribution of each spectral 
component to the total flux with the pulsed fraction of \psra\ in order to test
whether the pulsed flux could be ascribed to a specific thermal or non thermal 
component (or to a particular combination of them). 
For the case of the PL+BB model, the power law component contributes to 
$\sim$76\% of the pn counts in the 0.15--4 keV band, a value in perfect 
agreement with the pulsed  fraction of \psra\ ($75.7\pm5.4$)\% in the same 
energy range. However, the tentative association of most of the non thermal 
flux in the PL+BB model with the pulsed flux is ruled out by observing that, in 
this hypothesis, the steep power law should strongly contribute to the pulsed 
flux especially at low energies, but this is not the case; the power law 
contributes $\sim$86\% of the counts in the 0.15--0.3 keV band, a value much 
higher than the corresponding pulsed flux ($\sim$50\%) as shown by the stars 
plotted  in Figure~\ref{fig3}. A similar argument applies for the case of the 
double-blackbody model (see triangles plotted in Figure~\ref{fig3}). These 
results are confirmed by the phase-resolved spectra of \psra\ showing that the 
off-pulse spectrum cannot be well fitted by a single-component model.

More generally, none of the  double-component best-fit models are compatible
with the plot in Figure~\ref{fig3}, even with the assumption that the pulsed flux 
results from  a combination of both spectral components. In fact, the total 
flux of the softer spectral component (either the power law or the colder 
blackbody) is 3--4 times higher than that of the harder component. Therefore, 
it is impossible to invert the trend in Figure~\ref{fig3} even when assuming that the
harder component is fully pulsed. For example, in the case of the PL+BB model, 
assuming that the pulsed flux is due to the full blackbody component plus a 
part of the power law component, it is impossible to obtain a pulsed fraction that is
increasing with energy. Therefore, the entire X-ray flux of the double-pulsar 
system cannot be interpreted as a simple combination of a non thermal and 
single-temperature thermal emission from \psra\ or \psrb. Therefore, on the 
basis of these considerations and of the spectral results of \S ~\ref{spec}
we are led to consider models based on three spectral components.

The inclusion of an additional cooler blackbody BB$_{\rm{C}}$ (possibly 
related to \psrb) in the PL+BB model can account for the small pulsed fraction 
at low energy. Three-component models where blackbody emission (BB$_{\rm{H}}$) 
is responsible for most of the pulsed emission from \psra\ do not provide an 
acceptable spectral fit (see Table~\ref{tab_new}) and fail to match pulsed flux
fractions in Figure~\ref{fig3}. Assuming instead that most of the pulsed \psra\ 
flux is due to the power law, it is possible to reproduce the trend seen in 
Figure~\ref{fig3} (\emph{squares}). This is also the best three-component scenario 
compatible with the phase-resolved spectrum of \psra\ (see \S ~\ref{spec}, 
Table~\ref{tab_new}). It is worth noting that the phase-resolved spectrum of
\psra\ does not reveal any variation apart from a significant softening in the 
off-pulse phases, in agreement with the proposed three-component model.

The best three-component model fitting phase-resolved spectrum of \psra\ and
pulsed fluxes of both pulsars is thus composed of a power law with photon index
$\Gamma\sim3.3$ and 0.2--3 keV luminosity $L_{\rm{PL}}=1.9\times10^{30}$ \lum, 
a cooler ($\sim$30 eV) blackbody with bolometric luminosity 
$L_{\rm{BB_C}}=3.0\times10^{31}$ \lum, and a hotter blackbody ($\sim$130 eV)
with luminosity $L_{\rm{BB_H}}=5.2\times10^{29}$ \lum. The emission from \psra\
is mostly non thermal, its luminosity being equal to  $L_{\rm{PL}}$ plus the 
possible additional smaller contribution of $L_{\rm{BB_H}}$;
$L_{\rm{A}}\sim2\times10^{30}=3\times10^{-4}\dot{E}_{\rm{A}}$ in the 0.2--3 keV
energy range.

Because of the softness of the spectrum, the efficiency of the conversion of 
rotational energy loss into X-ray luminosity is strongly sensitive to the 
energy range considered. The \psra\ luminosity in the 0.1--10 keV range 
matches $\sim$$10^{-3}\dot{E}_{\rm{A}}$, the value typically found for
recycled radio pulsars that are detected at X-ray energies \citep{becker99}.
As expected from a mostly magnetospheric emission, pulse peaks are still 
relatively steep and far from thermal broad sinusoidal pulsations seen in
other recycled pulsars with comparable spin parameters \citep{zavlin02}. 
Non-thermal deeply modulated components are instead seen in the pulsed flux of 
the most energetic recycled pulsars (e.g. PSR\,B1821$-$24). The peculiar nature
of \psra\ is also stressed by the fact that no other recycled pulsar with such 
a steep non thermal emission component is known.

The observed equivalent emitting radius of BB$_{\rm{H}}$ is one order of 
magnitude smaller than the polar cap radius of \psra\ ($\sim$1 km), but
modeling with a non-uniform temperature of the heated region \citep{zavlin02} 
or invoking a partially filled polar cap \citep{hm02} could solve this 
discrepancy. Thus, we can interpret BB$_{\rm{H}}$ as due to the heating of A's
polar cap from back-flowing particles in its magnetosphere. The light curve of 
\psra\ shows a bridge between the two peaks 2--3 times higher than the 
off-pulse flux, leading to the idea that both non thermal pulses emerge from a 
wide, center-filled thermal emission cone from the magnetic pole possibly 
associated to BB$_{\rm{H}}$.

\subsection{The nature of \psrb\ X-ray emission}

We detected \psrb\ during part of the orbit, around the ascending node. Its 
0.2--3 keV flux  estimated from the pulsed light curve is 
$\sim$$1.2\times10^{-14}$ \flux\ and corresponds to ($35\pm15$)\% of the total flux 
in the orbital phase interval 0.44--0.61. The corresponding luminosity of 
$3.6\times10^{29}$ \lum\ represents $\sim$20\% of the spin-down energy loss, a 
value much higher than observed in all the other normal pulsars 
\citep{possenti02}.

The \psrb\ emission is likely associated to the cooler blackbody emission 
BB$_{\rm{C}}$ ($k_BT\simeq30$ eV), with the possible addition of the weaker 
and hotter component BB$_{\rm{H}}$ (if the latter is not associated to \psra).
In this case, the bolometric luminosity of \psrb\ can be calculated as
$L_{\rm{B}}\simeq4\times L_{\rm{BB_C}}\simeq1.2\times10^{32}$ \lum\
assuming it is emitted during one-fourth of the orbit ($L_{\rm{BB_C}}$ and 
$L_{\rm{BB_H}}$ are orbital phase-averaged luminosities).
This luminosity is surprisingly much higher than the X-ray emission from \psra\
(assuming $L_{\rm{A}}\simeq L_{\rm{PL}}=1.9\times10^{30}$ \lum). Obviously, 
X-ray emission from \psrb\ can only be powered by an external source, i.e. 
the spin-down energy from \psra. In this hypothesis, $\sim$2\% of the 
rotational energy loss of \psra\ is converted into thermal radiation from \psrb.

\citet{zhang04} explain the strong brightening of pulsed
radio flux from \psrb\ in two portions of the orbit as episodes
during which pairs from \psra's wind flow into the open field
line region of pulsar B and emit curvature radiation at radio
frequencies within an altitude of $\sim$$10^{8}$ cm. Once pairs
from A's wind leak into B's magnetosphere, they will stream all
the way down, heating the surface of B, giving rise to thermal
radiation. According to this model, the predicted X-ray luminosity
is $L_{\rm{X,B}}\sim2\times 10^{31}\eta\,(\Delta\Omega_{\rm{w,A}}/4\pi)^{-1}$ \lum,
and the typical temperature of the polar cap of \psrb\ is
$k_BT_{\rm{pc}}\sim0.4\,\eta^{1/4}(\Delta\Omega_{\rm{w,A}}/4\pi)^{-1/4}$ keV,
where $\eta$ is the fraction of all pairs from A's bow shock injected into B's 
open field line region and $\Delta\Omega_{\rm{w,A}}$ is the unknown solid angle
of A's wind. 

The X-ray luminosity predicted by the model is in good agreement with our 
results if we assume that \psra's wind is anisotropic (i.e. 
$\Delta\Omega_{\rm{w,A}}\lesssim1$ sr), as heuristically suggested by the 
variable illumination inducing radio and X-ray emission of \psrb. Within the 
above anisotropy assumption, the required efficiency $\eta$ is (poorly) 
constrained to $\leqslant$50\% to match the observed \psrb\ luminosity.

However, according to this model, \psrb's surface would be heated to a 
temperature ($\sim$500 eV) higher than the observed values of BB$_{\rm{C}}$ 
and BB$_{\rm{H}}$. This discrepancy could be due to the fact that most of the 
predicted luminosity (BB$_{\rm{C}}$) comes from a region larger than the polar 
cap and consistent with the whole neutron star's surface. The hotter component 
is consistent in size with the polar cap and could be responsible for the 
observed pulsations.


In radio, the shape and the intensity of the pulse profile of \psrb\ vary with
 orbital longitude, with two bright phases centered around longitudes (with 
respect to the ascending node) of $\sim$210$\degr$ (bright phase 1 [bp1], orbital phase 0.11) and 
280$\degr$ (bright phase 2 [bp2], orbital phase 0.31) at the epoch of first observations 
\citep{lyne04}. Because of the geodetic precession of pulsars' spins and periastron 
advance, these bright phases are shifting to greater longitudes at a rate of 
a few deg yr$^{-1}$ \citep{burgay05}. At the epoch of our \xmm\ observation, 
the centroid of the radio bright phase bp2 is expected at $\sim$300$\degr$ 
of longitude (orbital phase 0.36). The brightening of \psrb\ in X-rays appears 
in the orbital phase interval adjacent to (following) bp2 and peaks around the ascending 
node. When \psra's wind intercepts \psrb's magnetosphere, it powers radio 
emission and it starts heating B's surface. The X-ray emission from B lasts for
$30\pm10$ min each orbit (detection significance fades outside this range)
constraining the thermal inertial time of the neutron 
star in good agreement with theoretical expectations for 
external heating of a pulsar surface \citep{eichler89}.

\subsection{Constraints on X-ray emission from a bow shock between \psra's wind
and \psrb's magnetosphere}

The short occultation of A by B seen in the radio band
\citep{lyne04,mclaughlin04a,mclaughlin04b} could imply the presence of a dense
magnetosheath enfolding the magnetosphere of \psrb, although all the absorption could occur from within the \psrb's magnetosphere alone \citep{lt05}. In the magnetosheath model, the relativistic 
wind from A collides with B's magnetic field at an equilibrium distance from 
B comparable or smaller than the radius of its light cylinder 
\citep{arons05,lyutikov04}. Thus, in contrast with termination shocks seen in 
pulsar wind nebulae (PWNe) typically at distances $>$$10^6$ light cylinder 
radii, a termination shock between \psra\ and \psrb\ could probe the properties
of a pulsar's relativistic wind at a smaller distance from the central engine
than ever studied before. PWNe are usually strongly radiative at high energies 
with typical efficiencies $L_{\rm{X}}/\dot{E}_{\rm{PSR}}=10^{-2}$ to $10^{-3}$. 
Therefore, some trace of a peculiar PWN originating from \psrb's magnetic 
pressure confining the wind outflow from A should be in principle found as a 
non thermal component in our data.

An important issue concerning such a bow shock/magnetosheath model is that one 
might expect a flux modulation as a function of the orbital phase owing to the 
changing view of the shock front. In particular, the shocked wind is expected 
to flow away from the head of the bow shock in a direction roughly parallel to 
the shock \citep{lyutikov04,granot04}. This might imply relativistic beaming of
the radiation emitted by the shocked plasma, resulting in a $\lesssim$50$\%$ 
modulation of the observed emission as a function of the orbital phase with 
possible peaks at 90$\degr$ from conjunctions.

The upper limits we obtained on the flux variations as a function of the 
orbital phase allow us to exclude any significant orbital modulation 
$\geqslant$15\% in X-rays (\S ~\ref{timinga}). Assuming that most of the 
unpulsed flux of the system ($\sim$10\%--50\% of the total flux depending on the 
selected energy range) could be ascribed to the bow shock, flux variations 
should have been detected. In fact, the range of allowed spectral slopes of
non thermal components in our spectral fits is not compatible with the 
``canonical'' shock value $\Gamma\sim2$, although a soft post-shock spectrum 
could be explained with the presence of an unusual low-energy relativistic 
electron population, as required in the magnetosheath model to provide A and 
B's eclipses by synchrotron absorption \citep{arons05,lyutikov04}. 
Nevertheless, the off-pulse spectrum of \psra\ is not even compatible with a 
simple non thermal emission, and \psrb\ would fill most of it at least for the 
part of the orbit in which it is detected. Therefore there is not much room 
left for a shock emission component in our data. An upper limit (90\% confidence level) of 
$\sim$$10^{29}$ \lum\ (0.2--3 keV) corresponding to an efficiency of 
$L^{\rm{shock}}_{\rm{X}}/\dot{E}_{\rm{A}}=2\times10^{-5}$ can be obtained by 
evaluating the maximum allowed luminosity of an unpulsed power law component 
with photon index $\sim$2 in the off-pulse spectrum of \psra.

Such a low output in shock emission compared to typical PWN's 
$L^{\rm{shock}}_{\rm{X}}/ \dot{E}_{\rm{PSR}}$ is not surprising looking at the 
small solid angle over which the wind from A is intercepted by B and 
considering that the termination shock of the wind is much closer than in known
nebulae and that it could show a different phenomenology.

\begin{figure}
\resizebox{\hsize}{!}{\includegraphics[angle=00]{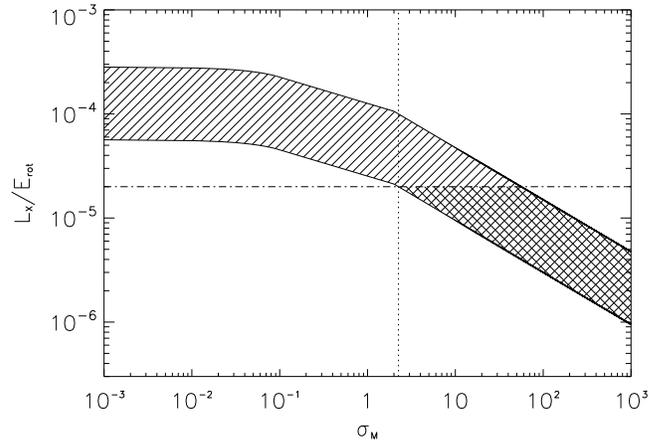}}
\caption{\label{sigma} Predicted X-ray emission efficiency
of a bow shock between \psra's wind and \psrb's magnetosphere
as a function of the wind magnetization parameter $\sigma_{M}$ (\emph{hatched
area}; see text).
The observed upper limit on the flux of a shock signature emission
in our data (\emph{dot-dashed line}) constrains $\sigma_{M}$ to values $>$1 (\emph{dotted line}).}
\end{figure}

Assuming that the wind energy is radiated isotropically from \psra\ and that it
is intercepted by a sphere centered on \psrb\ with radius equal to its light 
cylinder $R_{\rm{lc,B}}$, the resulting shock emission efficiency is 
$L^{\rm{shock}}_{\rm{X}}/ \dot{E}_{\rm{PSR}}=k R_{\rm{lc,B}}^2/4d^2 
 f(\sigma_M)$, where $d$ is pulsars' separation, $f(\sigma_M)$ is the 
fraction of power intercepted by the shock feeding into the accelerated 
electrons (it is a function of the unknown magnetization parameter $\sigma_M$),
and $k$ is a normalization factor to fit observed ``standard" PWNe X-ray 
efficiencies, where $\sigma_M\ll1$. \citet{kennel84,kc84} provide an extensive
discussion of the shock power function $f(\sigma_M)$ that is proportional to 
$1/(\sigma_M)^{1/2}$ for $\sigma_M\gg1$ (high-$\sigma_M$ shocks have poor 
efficiency) and is $\lesssim$1 for $\sigma_M\ll1$. Figure~\ref{sigma} shows the 
resulting $L^{\rm{shock}}_{\rm{X}}/\dot{E}_{\rm{PSR}}$ curve as a function of 
the magnetization parameter $\sigma_M$ accounting for uncertainties (\emph{hatched area}) associated with the actual size of the region intercepting the shock (it 
could be a factor 5 smaller than $R_{\rm{lc,B}}$ according to pressure balance)
and errors on $k$ due to the observed span in PWNe efficiencies. For low values
of the magnetization parameter, the shock could theoretically account for a 
significant fraction of the X-ray luminosity, but this is not the case; the 
observed upper limit (\emph{dot-dashed line}) constrains the magnetization parameter 
to values $\gtrsim$1 and, thus, much higher than those estimated for Crab-like
PWNe. Such an high magnetization parameter is apparently inconsistent with the 
value of $\sim$0.03 predicted by the magnetosheath model \citep{arons05}, and it
seems to confirm most of the modern wind models \citep{contopoulos02,lyubarsky01} 
predicting $\sigma_M\gg1$.

\citet{granot04} proposed a model in which the emission from the interaction of
the two pulsars is lower than that expected from the interaction of pulsar A's 
wind alone with the interstellar medium. Thus, \psra\ could provide a ``standard" PWN 
($\sigma_M\ll1$) at distances of $\sim$$10^{15}$ cm from the pulsar surface. In
this model, little or no modulation with the orbital periods $P_{\rm{A}}$ and 
$P_{\rm{B}}$ is expected, but even in this case it is difficult to account for 
the steepness of the unpulsed spectrum. Furthermore, the predicted luminosity 
of this model in the energy range (0.2--10 keV) is $\sim$$7\times10^{29}$ \lum,
much higher than our upper limit on the shock component discussed above.

\section{Conclusions}

The analysis of the $\sim$5600 source photons obtained from our $\sim$230 ks 
long \xmm\ observation confirmed the pulsed detection in X-ray of the \psra\ emission
providing 80\% of the total absorbed flux from the system. For the first time, 
X-ray emission from \psrb\ is also detected with good confidence ($\sim$200--300
pulsed counts) around the ascending node of the orbit.

A three-component spectral model can satisfactorily fit phase-resolved spectra 
and the energy dependence of the \psra\ pulsed fraction. The best-fit model is 
composed of a soft power law responsible for most of the pulsed emission from 
\psra\ (accounting for only $\sim$15\% of the total unabsorbed luminosity at 
$E>0.1$ keV), a cooler ($\sim$30 eV) blackbody likely associated to \psrb, and 
a hotter and fainter thermal component originating from back falling particles 
heating polar caps of either \psra\ or \psrb\ ($R_{\rm{BB_H}}\sim100$ m). Note
that the cooler blackbody accounts for most of the bolometric luminosity, even
if its contribution in the \xmm\ energy range is small.

\psra\ shows peculiar properties with respect to other known recycled pulsars 
(e.g. the softest non thermal spectrum), possibly
due to the particular evolutionary history of a double-neutron star system. The
phenomenology of \psrb\ is completely different from that of any other pulsar 
and the only viable possibility to explain its X-ray emission seems that it is 
powered by \psra's wind heating B's surface. Our observation is in good 
agreement with theoretical predictions by models also explaining B's radio 
flares \citep{zhang04}, although a more complex thermal emission scenario 
(colder and non uniform temperature) should be invoked.

There are no signs (e.g. orbital flux modulation, a significant non thermal
component in \psra's off-pulse) of the presence of X-ray emission from a 
bow shock between \psra's wind and \psrb's magnetosphere \citep{lyutikov04,arons05} 
invoked to explain the occultation of the radio emission of \psra\ at the inferior 
conjunction of \psrb. The upper limit on the flux of such a shock component 
constrains the wind magnetization parameter $\sigma_M$ of \psra\ to values 
$>$1, much higher than that predicted by the magnetosheath radio occultation 
of \psra. The absorption causing A's occultation is likely to occur within 
B's magnetosphere which retains enough plasma to produce an eclipse 
\citep{lt05}.

Further X-ray observations of \psr\ with \xmm\ and \cha\ with reasonable 
exposure times ($<$1 Ms) would not significantly improve \psrb\ detection or 
better constrain the pulsar's spectra. At gamma-ray energies, the absence of the 
strong thermal components seen in X-rays could instead allow one to reveal 
non thermal emission from the pulsars' interactions and better constrain the 
magnetospheric emission from \psra. The empirical relation between gamma-ray 
luminosity $L_{\gamma}$ and $\dot{E}$ for known gamma-ray pulsars \citep[i.e, 
$L_{\gamma}\propto (\dot{E})^{1/2}$,][]{thompson04,zh00} yields 
$L_{\gamma,\rm{A}}\sim10^{33}$ \lum, a value compatible with the luminosity of 
the possibly related EGRET source 3EG\,J0747$-$3412 \citep{hartman99} assuming
the same distance. The gamma-ray pulsed emission of \psra\ could be detected by
the \emph{AGILE} \citep{tavani06} and \emph{GLAST}\footnote{See \url{http://glast.gsfc.nasa.gov/}.} missions.

Apart from \psr, the only other known X-ray-emitting double-neutron star 
binary (DNSB) is PSR\,J1537+1155 \citep{kargaltsev06}. The X-ray spectra of 
PSR\,J1537+1155 and \psr\ are similar, and their X-ray luminosities are about
the same fraction of the respective spin-down energies. We note that, if  
the other DNSBs also have a similar efficiency, no other known object of this class 
can be detected by the currently available X-ray telescopes. Unlike 
\psr, in PSR\,J1537+1155 the distribution of photon arrival times over 
the binary orbital phase shows a deficit of X-ray emission around apastron, 
supporting the idea that most of the emission is caused by interaction of the 
relativistic wind from the pulsar with its neutron star companion. The 
difference between the two DNSBs can be explained by the smaller eccentricity 
of \psr\ or different alignments between the equatorial plane of the 
millisecond pulsar and the orbital plane of the binary \citep{kargaltsev06}.
Despite PSR\,J1537+1155 being weaker than \psr, it may offer better diagnostics
to discriminate emission produced by the shocked wind from ``standard" pulsar 
magnetospheric and surface emission, and thus, it is one of the most interesting 
systems suitable for further X-ray investigations of the inner pulsar 
magnetospheres and wind parameters in compact systems.

\acknowledgments
We thank Andrea Possenti for helpful discussions about PSR\,A timing analysis and cross-checks on timing corrections from our programs and Tempo.
The authors acknowledge the support of the Italian Space Agency (contract 
ASI/INAF I/023/05/0). A.D.L. acknowledges an Italian Space Agency fellowship.
 
\bibliographystyle{apj}
\bibliography{biblio}

\begin{thebibliography}{40}
\expandafter\ifx\csname natexlab\endcsname\relax\def\natexlab#1{#1}\fi

\bibitem[{{Anders} \& {Grevesse}(1989)}]{anders89}
{Anders}, E. \& {Grevesse}, N. 1989, \gca, 53, 197

\bibitem[{{Arons} {et~al.}(2005){Arons}, {Backer}, {Spitkovsky}, \&
  {Kaspi}}]{arons05}
{Arons}, J., {Backer}, D.~C., {Spitkovsky}, A., \& {Kaspi}, V.~M. 2005, in
  ASP Conf. Ser. 328, Binary Radio
  Pulsars, ed. F.~A. {Rasio} \& I.~H. {Stairs}, 95

\bibitem[{{Arons} \& {Tavani}(1993)}]{arons93}
{Arons}, J. \& {Tavani}, M. 1993, \apj, 403, 249

\bibitem[{{Balucinska-Church} \& {McCammon}(1992)}]{balucinska92}
{Balucinska-Church}, M. \& {McCammon}, D. 1992, \apj, 400, 699

\bibitem[{{Becker} \& {Tr{\"u}mper}(1999)}]{becker99}
{Becker}, W. \& {Tr{\"u}mper}, J. 1999, \aap, 341, 803

\bibitem[{{Blandford} \& {Teukolsky}(1976)}]{blandford76}
{Blandford}, R. \& {Teukolsky}, S.~A. 1976, \apj, 205, 580

\bibitem[{{Buccheri} {et~al.}(1983){Buccheri}, {Bennett}, {Bignami}, {Bloemen},
  {Boriakoff}, {Caraveo}, {Hermsen}, {Kanbach}, {Manchester}, {Masnou},
  {Mayer-Hasselwander}, {Ozel}, {Paul}, {Sacco}, {Scarsi}, \&
  {Strong}}]{buccheri83}
{Buccheri}, R., {Bennett}, K., {Bignami}, G.~F., {et~al.} 1983,
  \aap, 128, 245

\bibitem[{{Burgay} {et~al.}(2003){Burgay}, {D'Amico}, {Possenti}, {Manchester},
  {Lyne}, {Joshi}, {McLaughlin}, {Kramer}, {Sarkissian}, {Camilo}, {Kalogera},
  {Kim}, \& {Lorimer}}]{burgay03}
{Burgay}, M., {D'Amico}, N., {Possenti}, A., {et~al.} 2003,
  \nat, 426, 531

\bibitem[{{Burgay} {et~al.}(2005){Burgay}, {Possenti}, {Manchester}, {Kramer},
  {McLaughlin}, {Lorimer}, {Stairs}, {Joshi}, {Lyne}, {Camilo}, {D'Amico},
  {Freire}, {Sarkissian}, {Hotan}, \& {Hobbs}}]{burgay05}
{Burgay}, M., {Possenti}, A., {Manchester}, R.~N., {et~al.} 2005, \apjl, 624, L113

\bibitem[{{Campana} {et~al.}(2004){Campana}, {Possenti}, \&
  {Burgay}}]{campana04}
{Campana}, S., {Possenti}, A., \& {Burgay}, M. 2004, \apjl, 613, L53

\bibitem[{{Chatterjee} {et~al.}(2007){Chatterjee}, {Gaensler}, {Melatos},
  {Brisken}, \& {Stappers}}]{chatterjee07}
{Chatterjee}, S., {Gaensler}, B.~M., {Melatos}, A., {Brisken}, W.~F., \&
  {Stappers}, B.~W. 2007, \apj, 670, 1301

\bibitem[{{Contopoulos} \& {Kazanas}(2002)}]{contopoulos02}
{Contopoulos}, I. \& {Kazanas}, D. 2002, \apj, 566, 336

\bibitem[{{Cordes} \& {Lazio}(2002)}]{cordes02}
{Cordes}, J.~M. \& {Lazio}, T.~J.~W. 2002, preprint (astro-ph/0207156)

\bibitem[{{De Luca} \& {Molendi}(2004)}]{deluca04}
{De Luca}, A. \& {Molendi}, S. 2004, \aap, 419, 837

\bibitem[{{Eichler} \& {Cheng}(1989)}]{eichler89}
{Eichler}, D. \& {Cheng}, A.~F. 1989, \apj, 336, 360

\bibitem[{{Granot} \& {M{\'e}sz{\'a}ros}(2004)}]{granot04}
{Granot}, J. \& {M{\'e}sz{\'a}ros}, P. 2004, \apjl, 609, L17

\bibitem[{{Harding} \& {Muslimov}(2002)}]{hm02}
{Harding}, A.~K. \& {Muslimov}, A.~G. 2002, \apj, 568, 862

\bibitem[{{Hartman} {et~al.}(1999){Hartman}, {Bertsch}, {Bloom}, {Chen},
  {Deines-Jones}, {Esposito}, {Fichtel}, {Friedlander}, {Hunter}, {McDonald},
  {Sreekumar}, {Thompson}, {Jones}, {Lin}, {Michelson}, {Nolan}, {Tompkins},
  {Kanbach}, {Mayer-Hasselwander}, {M{\"u}cke}, {Pohl}, {Reimer}, {Kniffen},
  {Schneid}, {von Montigny}, {Mukherjee}, \& {Dingus}}]{hartman99}
{Hartman}, R.~C., {Bertsch}, D.~L., {Bloom}, {et~al.} 1999, \apjs, 123, 79

\bibitem[{{Kargaltsev} {et~al.}(2006){Kargaltsev}, {Pavlov}, \&
  {Garmire}}]{kargaltsev06}
{Kargaltsev}, O., {Pavlov}, G.~G., \& {Garmire}, G.~P. 2006, \apj, 646, 1139

\bibitem[{{Kennel} \& {Coroniti}(1984{\natexlab{a}})}]{kennel84}
{Kennel}, C.~F. \& {Coroniti}, F.~V. 1984{\natexlab{a}}, \apj, 283, 694

\bibitem[{{Kennel} \& {Coroniti}(1984{\natexlab{b}})}]{kc84}
---. 1984{\natexlab{b}}, \apj, 283, 710

\bibitem[{{Kramer} {et~al.}(2006){Kramer}, {Stairs}, {Manchester},
  {McLaughlin}, {Lyne}, {Ferdman}, {Burgay}, {Lorimer}, {Possenti}, {D'Amico},
  {Sarkissian}, {Hobbs}, {Reynolds}, {Freire}, \& {Camilo}}]{kramer06}
{Kramer}, M., {Stairs}, I.~H., {Manchester}, R.~N., {et~al.} 2006, Science, 314, 97

\bibitem[{{Lyne} {et~al.}(2004){Lyne}, {Burgay}, {Kramer}, {Possenti},
  {Manchester}, {Camilo}, {McLaughlin}, {Lorimer}, {D'Amico}, {Joshi},
  {Reynolds}, \& {Freire}}]{lyne04}
{Lyne}, A.~G., {Burgay}, M., {Kramer}, M., {et~al.} 2004, Science, 303, 1153

\bibitem[{{Lyubarsky} \& {Kirk}(2001)}]{lyubarsky01}
{Lyubarsky}, Y. \& {Kirk}, J.~G. 2001, \apj, 547, 437

\bibitem[{{Lyutikov}(2004)}]{lyutikov04}
{Lyutikov}, M. 2004, \mnras, 353, 1095

\bibitem[{{Lyutikov}(2005)}]{lyutikov05}
---. 2005, \mnras, 362, 1078

\bibitem[{{Lyutikov} \& {Thompson}(2005)}]{lt05}
{Lyutikov}, M. \& {Thompson}, C. 2005, \apj, 634, 1223

\bibitem[{{McLaughlin} {et~al.}(2004{\natexlab{a}}){McLaughlin}, {Camilo},
  {Burgay}, {D'Amico}, {Joshi}, {Kramer}, {Lorimer}, {Lyne}, {Manchester}, \&
  {Possenti}}]{mclaughlin04}
{McLaughlin}, M.~A., {Camilo}, F., {Burgay}, M., {et~al.} 2004{\natexlab{a}}, \apjl, 605, L41

\bibitem[{{McLaughlin} {et~al.}(2004{\natexlab{b}}){McLaughlin}, {Kramer},
  {Lyne}, {Lorimer}, {Stairs}, {Possenti}, {Manchester}, {Freire}, {Joshi},
  {Burgay}, {Camilo}, \& {D'Amico}}]{mclaughlin04b}
{McLaughlin}, M.~A., {Kramer}, M., {Lyne}, A.~G., {et~al.} 2004{\natexlab{b}},
  \apjl, 613, L57

\bibitem[{{McLaughlin} {et~al.}(2004{\natexlab{c}}){McLaughlin}, {Lyne},
  {Lorimer}, {Possenti}, {Manchester}, {Camilo}, {Stairs}, {Kramer}, {Burgay},
  {D'Amico}, {Freire}, {Joshi}, \& {Bhat}}]{mclaughlin04a}
{McLaughlin}, M.~A., {Lyne}, A.~G., {Lorimer}, D.~R., {et~al.}
  2004{\natexlab{c}}, \apjl, 616, L131

\bibitem[{{Miller} {et~al.}(2003){Miller}, {Wijnands}, {M{\'e}ndez},
  {Kendziorra}, {Tiengo}, {van der Klis}, {Chakrabarty}, {Gaensler}, \&
  {Lewin}}]{miller03}
{Miller}, J.~M., {Wijnands}, R., {M{\'e}ndez}, M., {et~al.} 2003, \apjl, 583, L99

\bibitem[{{Pellizzoni} {et~al.}(2004){Pellizzoni}, {De Luca}, {Mereghetti},
  {Tiengo}, {Mattana}, {Caraveo}, {Tavani}, \& {Bignami}}]{pellizzoni04}
{Pellizzoni}, A., {De Luca}, A., {Mereghetti}, S., {Tiengo}, A., {Mattana}, F.,
  {Caraveo}, P., {Tavani}, M., \& {Bignami}, G.~F. 2004, \apjl, 612, L49

\bibitem[{{Possenti} {et~al.}(2002){Possenti}, {Cerutti}, {Colpi}, \&
  {Mereghetti}}]{possenti02}
{Possenti}, A., {Cerutti}, R., {Colpi}, M., \& {Mereghetti}, S. 2002, \aap,
  387, 993

\bibitem[{{Str{\"u}der} {et~al.}(2001){Str{\"u}der}, {Briel}, {Dennerl},
  {Hartmann}, {Kendziorra}, {Meidinger}, {Pfeffermann}, {Reppin}, {Aschenbach},
  {Bornemann}, {Br{\"a}uninger}, {Burkert}, {Elender}, {Freyberg}, {Haberl},
  {Hartner}, {Heuschmann}, {Hippmann}, {Kastelic}, {Kemmer}, {Kettenring},
  {Kink}, {Krause}, {M{\"u}ller}, {Oppitz}, {Pietsch}, {Popp}, {Predehl},
  {Read}, {Stephan}, {St{\"o}tter}, {Tr{\"u}mper}, {Holl}, {Kemmer}, {Soltau},
  {St{\"o}tter}, {Weber}, {Weichert}, {von Zanthier}, {Carathanassis}, {Lutz},
  {Richter}, {Solc}, {B{\"o}ttcher}, {Kuster}, {Staubert}, {Abbey}, {Holland},
  {Turner}, {Balasini}, {Bignami}, {La Palombara}, {Villa}, {Buttler},
  {Gianini}, {Lain{\'e}}, {Lumb}, \& {Dhez}}]{struder01}
{Str{\"u}der}, L., {Briel}, U., {Dennerl}, K., {et~al.} 2001, \aap, 365, L18

\bibitem[{{Tavani} {et~al.}(2006){Tavani}, {Barbiellini}, {Argan}, {Basset},
  {Boffelli}, {Bulgarelli}, {Caraveo}, {Chen}, {Costa}, {De Paris}, {Del
  Monte}, {Di Cocco}, {Donnarumma}, {Feroci}, {Fiorini}, {Foggetta},
  {Froysland}, {Frutti}, {Fuschino}, {Galli}, {Gianotti}, {Giuliani},
  {Labanti}, {Lapshov}, {Lazzarotto}, {Liello}, {Lipari}, {Longo}, {Marisaldi},
  {Mastropietro}, {Mattaini}, {Mauri}, {Mereghetti}, {Morelli}, {Morselli},
  {Pacciani}, {Pellizzoni}, {Perotti}, {Picozza}, {Pittori}, {Pontoni},
  {Porrovecchio}, {Prest}, {Pucella}, {Rapisarda}, {Rossi}, {Rubini},
  {Soffitta}, {Traci}, {Trifoglio}, {Trois}, {Vallazza}, {Vercellone},
  {Zambra}, \& {Zanello}}]{tavani06}
{Tavani}, M., {et~al.} 2006, Proc. SPIE, 6266, 2

\bibitem[{{Thompson}(2004)}]{thompson04}
{Thompson}, D.~J. 2004, in Astrophysics and Space Science Library, Vol. 304,
  Cosmic Gamma-Ray Sources, ed. K.~S. {Cheng} \& G.~E. {Romero}, 149

\bibitem[{{Turner} {et~al.}(2001){Turner}, {Abbey}, {Arnaud}, {Balasini},
  {Barbera}, {Belsole}, {Bennie}, {Bernard}, {Bignami}, {Boer}, {Briel},
  {Butler}, {Cara}, {Chabaud}, {Cole}, {Collura}, {Conte}, {Cros}, {Denby},
  {Dhez}, {Di Coco}, {Dowson}, {Ferrando}, {Ghizzardi}, {Gianotti}, {Goodall},
  {Gretton}, {Griffiths}, {Hainaut}, {Hochedez}, {Holland}, {Jourdain},
  {Kendziorra}, {Lagostina}, {Laine}, {La Palombara}, {Lortholary}, {Lumb},
  {Marty}, {Molendi}, {Pigot}, {Poindron}, {Pounds}, {Reeves}, {Reppin},
  {Rothenflug}, {Salvetat}, {Sauvageot}, {Schmitt}, {Sembay}, {Short},
  {Spragg}, {Stephen}, {Str{\"u}der}, {Tiengo}, {Trifoglio}, {Tr{\"u}mper},
  {Vercellone}, {Vigroux}, {Villa}, {Ward}, {Whitehead}, \& {Zonca}}]{turner01}
{Turner}, M.~J.~L., {Abbey}, A., {Arnaud}, M., {et~al.}
  2001, \aap, 365, L27

\bibitem[{{Zavlin} {et~al.}(2002){Zavlin}, {Pavlov}, {Sanwal}, {Manchester},
  {Tr{\"u}mper}, {Halpern}, \& {Becker}}]{zavlin02}
{Zavlin}, V.~E., {Pavlov}, G.~G., {Sanwal}, D., {Manchester}, R.~N.,
  {Tr{\"u}mper}, J., {Halpern}, J.~P., \& {Becker}, W. 2002, \apj, 569, 894

\bibitem[{{Zhang} \& {Harding}(2000)}]{zh00}
{Zhang}, B. \& {Harding}, A.~K. 2000, \apj, 532, 1150

\bibitem[{{Zhang} \& {Loeb}(2004)}]{zhang04}
{Zhang}, B. \& {Loeb}, A. 2004, \apjl, 614, L53

\end{thebibliography}

\end{document}